\documentclass{llncs}

\usepackage[lined,boxed,commentsnumbered,ruled,vlined,noend,
boxed]{algorithm2e}

\usepackage{amsfonts}
\usepackage{color}

\newcommand{\remove}[1]{}
\newtheorem{result}{Result}
\newtheorem{invariant}{Invariant}

\newcommand{\IR}{\mathbb{R}}
\newcommand{\CB}{\textsf{\sc Compute-Bridge}}
\newcommand{\KUH}{\textsf{\sc KS-Upper-Hull}}
\newcommand{\ROUH}{\textsf{\sc Read-Only-Upper-Hull}}
\newcommand{\MDLP}{\textsf{\sc Megiddo's-2D-LP}}
\newcommand{\MDLPT}{\textsf{\sc Megiddo's-3D-LP}}
\newcommand{\TL}{\textsf{\sc Testing-Line}}

\title{Convex Hull and Linear Programming in Read-only Setup with Limited
Work-space}
\author{Minati De\inst{1} \and Subhas C. Nandy\inst{1} \and Sasanka Roy\inst{2}}
\institute{Indian Statistical Institute, Kolkata - 700108, India \and Chennai
Mathematical Institute, Chennai - 603103, India}

\begin{document}

\maketitle

\vspace{-0.2in}
\begin{abstract} 
{\it Prune-and-search} is an important paradigm for solving many important
geometric problems. We show that the general {\it prune-and-search} technique
can be implemented where the objects are given in read-only memory. As examples
we consider convex-hull in 2D, and linear programming in 2D and 3D. For
the convex-hull problem, designing sub-quadratic algorithm in a read-only
setup with sub-linear space is an open problem for a long time. We first
propose a simple algorithm for this problem that runs in
$O(n^{\frac{3}{2}+\epsilon)}$ time and $O(n^\frac{1}{2})$ space. Next, we
consider a restricted version of the problem where the points in $P$ are given
in sorted order with respect to their $x$-coordinates in a read-only array. For
the linear programming problems, the constraints are given in  the read-only
array. The last three algorithms use {\it prune-and-search}, and their time and
extra work-space complexities are $O(n^{1 + \epsilon})$ and $O(\log n)$
respectively, where $\epsilon$ is a small constant satisfying
$\sqrt{\frac{\log\log n}{\log n}} < \epsilon < 1$. 
\vspace{-0.2in}
\end{abstract}

\section{Introduction}
\vspace{-0.1in}
Designing algorithmic tools with fast and limited size memory (e.g caches) but
having  capability of very fast processing of a massive high quality data is a
challenging field of research \cite{AsanoMRW11,BV,BMM07}. The problem becomes
much more difficult if the input data is given in a read-only array, and very
small amount of work-space
is available in the system. Such a situation arises in the concurrent
programming environment where many processes access the same data, and hence
modifying the data by a process during the execution is not permissible
\cite{barba}. In this paper, we show that the general {\it prune-and-search}
technique can be implemented where the objects are given in read-only array. As
examples we consider convex-hull in 2D, and linear programming in 2D and
3D.

Given a set $P=\{p_1,p_2,\ldots,p_n\}$ of points in 2D, the
problem of designing sub-quadratic time algorithm for computing convex hull for
$P$
with sub-linear extra work-space is an important problem and is being studied
for a long time. Bronnimann et al. \cite{BronnimannIKMMT02} showed that Graham's
scan algorithm for computing convex hull of a planar point set of size $n$ can
be made in-place maintaining $O(n\log n)$ time complexity. Here, the extra
workspace required is $O(1)$, and the output is available in the same input array.
In the same paper they also showed that (i) the convex hull of a point set  in
2D can be computed in an in-place manner in $O(n\log h)$ time and
with $O(1)$ extra workspace where $h$ is the number of hull vertices, and (ii) 
the linear programming in 2D with $n$ constraints can be solved in
$O(n)$ time. Very recently, Vahrenhold \cite{Vahrenhold12} showed that the
prune-and-search algorithm by Kirkpatrick and Seidel \cite{KirkpatrickS86} for
computing the convex hull of a planar point set can also be made in-place
maintaining the $O(n\log h)$ time complexity and using only $O(1)$ work-space.
All these algorithms permute the input array after the execution. If a planar
point set $P$ is given in a read-only array, then the well-known Jarvis March
algorithm computes the convex hull in $O(nh)$ time with $O(1)$ extra space. The
problem of designing a sub-quadratic algorithm for computing convex hull in a
read-only environment with sub-linear work-space is an open problem for a long
time. Chan and Chen \cite{ChanC07} proposed an algorithm that can compute the
convex hull in $O(n(\log n+\frac{n}{s}))$ time using $O(s)$ space where $s \leq
n$ is a chosen integer. In the same paper, they proposed an $O(n)$ time
randomized algorithm for the linear programming problem in fixed dimension using
$O(\log n)$ extra space in a read-only environment. They also considered the
problem of computing the convex hull where the points are sorted by their
$x$-coordinates. The proposed algorithm is a randomized one and runs in
$O(\frac{1}{\delta}n)$ expected time and $O(\frac{1}{\delta}n^\delta)$ extra
space for any fixed $\delta >0$. The algorithm can be made deterministic if the
running time is increased to $O(2^{O(\frac{1}{\delta})}n)$. The convex hull of
a simple polygon with $n$ vertices can be computed in a read-only setup in
$O(\frac{n\log n}{\log p})$ time with $O(\frac{p\log n}{\log p})$ extra
workspace \cite{barba}.

In this paper, we first address the open problem related to the convex hull
problem in 2D. We show that if the points in $P$ are given in a read-only array
then the convex hull can be computed in $O(n^{\frac{3}{2}+\epsilon)}$ time and
$O(n^\frac{1}{2})$ extra space. Next, we consider a restricted version of the
convex hull problem, where the input points are given in sorted order of their
$x$-coordinates. Here, we can apply {\it prune-and-search} technique to compute
the convex hull of $P$ in $O(n^{1+\epsilon})$ time and $O(\log n)$ space, where
$\epsilon$ is a constant satisfying $\sqrt{\frac{\log\log n}{\log n}} < \epsilon
< 1$. We also show that similar technique works for solving the linear
programming problem in 2D and 3D in the read-only setup with the same time
complexity. In this context, it needs to be mentioned that a similar technique
is adopted to solve the minimum enclosing circle problem for a set of points in
2D, where the input points are given in a read-only array \cite{D12}. 

\vspace{-0.2in}
\section{Convex hull}
\vspace{-0.2in}
\subsection{Unrestricted version}\vspace{-0.1in}
Given a set $P$ of $n$ points in 2D in a read-only array, the objective is to
report the vertices of the convex hull of $P$. We describe the method of
reporting the upper-hull; the lower-hull can be computed in a similar manner.
We use three arrays, namely $A$, $B$ and $C$, each of size $O(\sqrt{n})$ as the
work-space. For the notational simplicity, we will use $P_{(i)}$ and $P_i$ to
denote the set of points whose $x$-coordinate lies between $x_{(i\sqrt{n}+1)}$
to $x_{((i+1)\sqrt{n})}$, and the set of points whose
$x$-coordinate is less than $x_{((i+1)\sqrt{n})}$ respectively, where $x_{(k)}$
denotes the $k$-th smallest element among the $x$-coordinates of the points in
$P$. The upper hull of $P_{(i)}$ and $P_i$ are denoted by $CH_{(i)}$ and $CH_i$
respectively. Our algorithm executes in two passes. Each pass consists of
$\lceil\sqrt{n}\rceil$ stages. In the $i$-th stage of the first pass, we pick up
the points in $P_{(i)}$ in the array $A$. We assume that $(i-1)$ stages are
complete; the vertices of $CH_{(i-1)}$ in the convex hull
$CH_{i-1}$ are stored in the array $B$. The $j$-th element of the array
$C$ (denoted by $C[j]$) contains the first and last hull-vertices $(f_j,\ell_j)$
among the points in $P_{(j)}$ in the convex hull $CH_{i-1}$, $j \leq i-1$. If
no such hull-vertex exists then $C[j]$ contains
$(-1,-1)$. We execute the following steps in the $i$-th stage.
\vspace{-0.1in}
\begin{itemize}
\item[1.] Compute $x_{(i\sqrt{n}+1)}$ and $x_{((i+1)\sqrt{n})}$ among the points
in the array $P$, and identify all the points in $P_{(i)}$ to store them in the
array $A$.
\item[2.] Compute the upper hull $CH_{(i)}$ of the points in $A$ using the
in-place convex hull algorithm of \cite{BronnimannIKMMT02}.
\item[3.] Merge $CH_{(i)}$ with $CH_{i-1}$ as follows: (i) draw the common
tangent $L=[a,b]$ of $CH_{(i-1)}$ (stored in $B$) and $CH_{(i)}$ , where $a
\in CH_{(i-1)}$ and $b \in CH_{(i)}$. If $a$ is not the first vertex of
$CH_{(i-1)}$, then update $C[i-1]$ by $(f_{i-1},a)$ and put $[b,\ell_i]$ in
$C[i]$ ($\ell_i$ is obtained from $A$). Otherwise (i.e., if $a$ is the
first vertex of $CH_{(i-1)}$) then traverse the array $C$ to identify a
hull-vertex $u \in CH_{(j)}$ of a preceding block $j$ ($j < i-1$) that is
connected with $a \in CH{(i-1)}$. Note that, if $j \leq i-2$, then all the array
elements $C[k]$, $j+1 \leq k \leq i-1$ will contain $(-1,-1)$. We recompute
$CH_{(j)}$ and draw the common tangent of $CH_{(i)}$ and $CH_{(j)}$. The same is
followed until we get a tangent of $CH_{(i)}$ and $CH_{(j')}$ ($j' < i-1$) that
does not touch the vertex $f_{j'}$. We update $C[j']$ and set $C[i]$ with
appropriate vertex pair. 
\end{itemize}
\vspace{-0.1in}
In the second pass, we compute $CH_{(i)}$ for all the blocks whose $C[i] \neq
(-1,-1)$, and report only the portion from $f_i$ to $\ell_i$.
\vspace{-0.1in}
\begin{theorem}

Given a set $P$ of $n$ points in 2D in a read-only array, the
convex-hull of $P$ can be correctly computed in $O(n^{\frac{3}{2}+\epsilon})$
time using $O(n^{\frac{1}{2}})$ extra-space, where  $\sqrt{\frac{\log\log
n}{\log n}} < \epsilon <1$. 
\end{theorem}
\vspace{-0.15in}
\begin{proof}
The correctness follows from the fact that in the $i$-th stage,
$CH_{(i)}$ is appropriately merged with $CH_{i-1}$. 
We now analyze the time complexity of the first pass. 
In each stage $i$, $x_{(i\sqrt{n}+1)}$ to $x_{((i+1)\sqrt{n})}$ can be computed
in $O(n^{1+\epsilon})$ time using $O(\frac{1}{\epsilon})$ extra space, where
$\sqrt{\frac{\log \log n}{\log n}}< \epsilon < 1$ (see
the algorithm of \cite{MunroR96} in Appendix 1). Next, the convex hull
$CH_{(i)}$ in the array $A$ is computed in $O(\sqrt{n}\log n)$ time
\cite{BronnimannIKMMT02}. While merging $CH_{(i)}$ with $CH_{i-1}$, we may need
to recompute $CH_{(j)}$ for different $j < i-1$. However, the
recomputation of the convex hull of a block implies that there exists a block
whose no vertex participate in the convex hull of $P$. Thus the amortized
complexity of pass 1 is $O(n^{\frac{3}{2}+\epsilon}+n \log n)$. The second
pass needs the same amount of time. The space complexity follows from the size
of $A$, $B$ and $C$, and the fact that $\frac{\log n}{\log \log n} < n$.
\qed
\end{proof}
\vspace{-0.2in}
\subsection{Restricted version}
\vspace{-0.15in}

Given a set $P$ of $n$ points in 2D sorted with respect to their
$x$-coordinates in a read-only array, the objective is to report the edges of
the convex hull of the points in $P$. We will show how  Kirkpatrick
and Seidel's~\cite{KirkpatrickS86} deterministic prune-and-search algorithm for
computing  convex hull can be implemented in this framework. 

The algorithm in \cite{KirkpatrickS86} computes upper-hull and lower-hull
separately and report them. The basic steps of computing upper-hull for a set of
points $P$  is given in Algorithm~\ref{KUH}. Lower-hull can be computed in a
similar  way. Algorithm~\ref{KUH} follows divide-and-conquer paradigm. It uses a
procedure \CB\ to compute the bridge between two disjoint subsets of $P$ using
the prune-and-search technique. The details of this  procedure is described in
Algorithm~\ref{CB}. 

\vspace{-0.2in}
\begin{algorithm}
 \small
 \caption{\KUH($P$)}
 \KwIn{A set of points $P$ in 2D sorted according to $x$-coordinates}
\KwOut{The upper-hull of $P$}
*(Uses divide-and-conquer technique)*\\
{\bf STEP 1:} Find the point $p_m \in P$ having median $x$-coordinate\;
 
{\bf STEP 2:} Partition $P$ into two subset $P_\ell$ and $P_r$ where
$P_\ell$ contains all the points in $P$ whose $x$-coordinate is less than or
equal to $x(p_m)$ and  $P_r= P\setminus P_\ell$\;

 {\bf STEP 3:} $(a,b)$=\CB($P_\ell,P_r$); (* This procedure 
computes the bridge between $P_\ell$ and $P_r$; $a \in P_\ell$, $b \in P_r$ *)
\\
{\bf STEP 4:} Report $(a,b)$\;

 {\bf STEP 5:} Compute $P_\ell=P_\ell \setminus {P'_\ell}$, where $P'_\ell=\{p
\in
P_\ell| x(p)>x(a)\}$, and\\
$P_r=P_r \setminus {P'_r}$, where $P'_r=\{p \in P_r|
x(p)<x(b)\}$\;

 {\bf STEP 6:} \KUH($P_\ell$)\;
{\bf STEP 7:} \KUH($P_r$)\;
 
\normalsize
\label{KUH}
\end{algorithm}

The straight-forward implementation of the algorithm \KUH\ in a read-only
memory
requires $O(n)$ space for the procedure \CB\, as it needs to remember which
points were pruned in the previous iterations. In addition, the algorithm 
\KUH($P$)  reports the hull-edges in an arbitrary fashion (not in order
along the boundary of the convex hull) and takes $O(\log n)$ space for the
recursions.  So, the main hurdle in read-only model is to (i) report the hull
edges in order,  and (ii) implement the procedure
\CB\ using only $O(\log n)$ extra-space. In the next subsections, we 
describe how to resolve these issues. With this we have following main result:

\begin{theorem} \label{th}
 Given a set of $n$ sorted points $P$ of 2D in a read-only array, the
convex-hull of $P$ can be computed in $O(n^{1+\epsilon})$ time using $O(\log n)$
extra-space, where  $\sqrt{\frac{\log\log n}{\log n}} < \epsilon <1$. 
\end{theorem}
\vspace{-0.3in}

\begin{algorithm}[t]
 \small
 \caption{\CB($P_\ell,P_r$)}
 \KwIn{Two sets of points in 2D, $P_\ell$ and $P_r$ sorted according to
$x$-coordinates}
\KwOut{The bridge between $P_\ell$ and $P_r$}
*(Uses prune-and-search technique)*\\
 {\bf STEP 1:}\\
 \While{$|P_\ell| > 1$ and $|P_r| > 1$}
{
    {\bf STEP 1.1:} Arbitrarily pair-up points in  $P_\ell \cup P_r$; Let $L$
    be the set of these pairs. Each such pair of points $(p,q) \in L$
    will signify a line $\overline{pq}$ which will pass through $p,q$. We denote
    the slope of the line $\overline{pq}$ as $\alpha(\overline{pq})$.

    {\bf STEP 1.2:} Consider the slopes of these $\frac{|P_\ell \cup
    P_r|}{2}$ lines and compute their median. Let $\alpha_m$ be the median
    slope.

    {\bf STEP 1.3:} Compute the supporting line of $P_\ell$ and $P_r$ with
    slope $\alpha_m$; Suppose these are at points $a$($\in P_\ell$) and $b$($\in
    P_r$) respectively.

    {\bf STEP 1.4:} Now compare $\alpha(\overline{ab})$ with  $\alpha_m$. Here
    one of the three cases may arise: (i) $\alpha(\overline{ab})=\alpha_m$,
    (ii) $\alpha(\overline{ab}) < \alpha_m$ or (iii)
    $\alpha(\overline{ab})>\alpha_m$. \\
    \If{$\alpha(\overline{ab})=\alpha_m$}{ $\overline{ab}$ is the
    required bridge between the points in
    $P_\ell$ and $P_r$. So, the procedure returns $(a,b)$.}
    {\bf Otherwise}\\ Decide whether $\alpha(\overline{ab}) < \alpha_m$ or
    $\alpha(\overline{ab}) > \alpha_m$.
    Without loss of generality, assume that $\alpha(\overline{ab})< \alpha_m$,
    then  we will consider all the pairs $(p,q) \in L$ whose
    $\alpha(\overline{pq})  \geq  \alpha_m$. We can ignore the one point among
    $(p,q)$ which is to the left  of the  other one. So, at least $\frac{|P_\ell
    \cup P_r|}{4}$  points are  ignored for  further consideration.\\

}

{\bf STEP 2:}\\
Find the bridge in brute-force manner and return  the bridge.

\normalsize
\label{CB}
\end{algorithm}

\subsubsection{Reporting the hull-edges in-order}
\vspace{-0.1in}

Now, we will show how to report the hull vertices in clock-wise order using no
more than $O(\log n)$ extra-space. Consider the recursion tree $\cal T$ of the
algorithm \KUH. Its each node represents the reporting of a hull-edge. In the
algorithm \KUH,  as the reporting is done according to pre-order traversal of
the tree $\cal T$, the hull edges are not reported in clock-wise order. In order
to report them in clock-wise order, we need to traverse the recursion tree in
in-order manner, i.e. STEP 3 and STEP 4 of the Algorithm \ref{KUH} should be in
between STEP 5 and STEP 6. But, if we do this, then we can not evoke
\KUH($P_\ell$) on the updated set $P_\ell$. To resolve this problem, we will
compute the bridge in STEP 3 itself but will not report it then. We push it in
the stack in STEP 3 and pop it from stack in between STEP 5 and STEP
6. The size of this stack depends on the depth of the recursion tree which is 
$O(\log h)$, where $h$ is the number of hull-edges. The details of this change
is given as \ROUH($start,end$) in Algorithm~\ref{ROUH}. Thus we have the
following result:

\begin{algorithm}[t]
\small
\caption{\ROUH($start,end$)}
\KwIn{A portion of read-only array $P[start, \ldots, end]$ containing points in 
2D sorted according to the $x$-coordinates }
\KwOut{The upper-hull for the points in  $P[start, \ldots, end]$}
{\bf STEP 1:} Let $m=\lceil\frac{end-start}{2}\rceil$; The point $P[m] \in
P[start, \ldots, end ]$ have the median $x$-coordinate\;
{\bf STEP 2:} (* Now $P_\ell=P[start,  \ldots, m ]$ and $P_r=P[m+1, \ldots,
end]$ *)\\
{\bf STEP 3:} $(i,j)$=\CB($start,m,end$); (* This procedure 
returns a pair of indices of points in  array $P$ that defines the bridge
between $P_\ell$ \& $P_r$ *),  \\
Push the edge $(P[i],P[j])$ in STACK\;
{\bf STEP 4:} (* Modified $P_\ell=P[start, \ldots, i]$ and $P_r=P[j, \ldots,
end]$ *)\\
{\bf STEP 5:} \ROUH($start,i]$)\;
{\bf STEP 6:}   Report the edge $(P[i],P[j])$ popping the top element from
STACK\;
{\bf STEP 7:} \ROUH($j,end]$)\;
\normalsize
\label{ROUH}
\end{algorithm}

\vspace{-0.1in}
\begin{lemma}
 Given a set of $n$ sorted points $P$ in a read-only array, the reporting of
the hull edges can be done in clock-wise order using only $O(\log n)$
extra-space (assuming that the procedure \CB\ takes no more
than $O(\log n)$
space). 
\end{lemma}
\vspace{-0.1in}

\vspace{-0.2in}
\subsubsection{\CB\ with $O(\log n)$ extra-space}
\vspace{-0.1in}
Here the input set of points $P$ of this procedure is first partitioned into
two parts $P_\ell$ and $P_r$ by choosing the point $P[m]$ having median
$x$-coordinate. Since the array is sorted with respect to the $x$-coordinates,
this needs $O(1)$ time. Now an iterative procedure (while-loop) is executed to
compute the bridge of $P_\ell$ and $P_r$. In each iteration of the while-loop of
the procedure \CB, $\frac{1}{4}$th of the points from the set $P_\ell \cup P_r$
are pruned. The points which are pruned in $i$th iteration are not considered in
any $j$th iteration, where $j >i$. After an iteration of the while-loop, either
the bridge is returned or the iteration continues until $|P_\ell|=1$ or
$|P_r|=1$. So, the number of iterations of the while loop is $O(\log n)$, where
$n$ is the total number of points in $P$.

While executing the $i$th iteration, we want to remember the points which were
pruned in the previous $i-1$ iterations, $i \in \{1, \ldots \log n\}$. If we use
mark-bits
to remember which points are valid/invalid, then we need $O(n)$ bits. But, we
have only $O(\log n)$ work-space to be used. So, we take an array $M$ of size
$O(\log n)$ and another bit-array $B$ of size $O(\log n)$. At each $i$th
iteration, ignoring all the pruned points, we pair the valid points and 
consider the slopes of all the lines defined by the paired points. We compute
the
median slope value $\mu_i$ of these lines and store it at $M[i]$.  If the
supporting lines are
at points $a$ and $b$, we set $B[i]$ as 1 or 0 depending on whether the slope
$\alpha(\overline{ab})$ is greater than or less than  $\mu_i$ (since
$\alpha(\overline{ab})=\mu_i$ implies that we already get the bridge). Thus,
$B[i]$ signifies whether $M[i]$ is greater than or less than the slope
$\alpha^*$ of
the desired bridge. 

Now, we will describe a pairing scheme which will satisfy the following
invariants:
\vspace{-0.2in}
\begin{invariant}
\begin{enumerate}
\item[(i)]  If a point $p$ is pruned at some iteration $i$, then it will not
participate to form a pair for any $j$-th iteration, where $j>i$.
\item[(ii)] If $(p,q)$ are paired at the $i$-th iteration of the while-loop, and
none of the points $p,q$ is pruned at the end of this iteration and we need to
go for $i+1$-th iteration, then $(p,q)$ will again form a valid pair at $i+1$-th
iteration.
\item[(iii)] If $(p,q)$ is a valid pair at $i$-th iteration and
$(p,s)$( $s \neq q$)  are paired at $i+1$-th iteration of the while-loop, then 
there exist some $r$ such that $(r,s)$ were paired at $i$-th iteration, and $q$
and $r$ were pruned at the end of $i$-th iteration.

\end{enumerate}
\label{inv1}
 \end{invariant}
\vspace{-0.1in}
 
The iteration starts with the points $\{P[start], P[start+1], \ldots,$
$P[end]\}$.
 In the first iteration of the while-loop, we consider the consecutive points, i.e,
$(P[start],$ $P[start+1])$, $(P[start+2], P[start+3])$, $\ldots$ as valid pairs.
 
Assume that first $i-1$ iterations of the while-loop are over, and we are at the
beginning of the $i$-th iteration;  $M[t]$ contains 
median slope of $t$-th iteration and 
$B[t]$ contains 0 or 1 depending on $M[t]> \alpha^*$ or $M[t]< \alpha^*$ for
all $1\leq t \leq i-1 $. Now, we want to detect all the valid points and pair
them up maintaining the Invariant~\ref{inv1}. We use another array $IndexP$ of
size $O(\log n)$ whose all elements are set to -1 at the beginning of this
iteration. 

We consider the point-pairs $(P[start+2\nu],P[start+2\nu+1])$, $\nu=0,1,\ldots,
\lfloor\frac{end-start+1}{2}\rfloor$ in order. For each pair, we compute the
slope
$\gamma=\alpha(\overline{P[start+2\nu],P[start+2\nu+1]})$ of the corresponding
line, and perform the level 1 test using $M[1]$ and $B[1]$ to see
whether both of them remain {\it valid} at iteration 1. If the
test succeeds, we perform level 2 test for $\gamma$ by using $M[2]$ and
$B[2]$. We proceed similarly until (i) we reach up to $i-1$-th level and both
the
points remain {\it valid} at all the levels, or (ii) one of these points becomes
{\it invalid} at some level, say $j$ ($< i-1$). In Case (i), the point
pair $(P[start+2\nu],P[start+2\nu+1])$ will form a valid pair and  participates
in computing the median value
$m_i$. In case (ii), suppose $P[start+2\nu]$ remains {\it valid} and
$P[start+2\nu+1]$
becomes {\it invalid}. Here two situations need to be considered depending on
the value of $IndexP[j]$. If $IndexP[j]=-1$ (no point is stored in
$IndexP[j]$), we
store $start+2\nu$ or $start+2\nu+1$ in $IndexP[j]$ depending on whether
$P[start+2\nu]$
or $P[start+2\nu+1]$ remains {\it valid} at level $j$. If $IndexP[j]=\beta (\neq
-1)$ (index of a {\it valid} point), we form a pair $(P[start+2\nu],P[\beta])$
and
proceed to check starting from $j+1$-th level (i.e., using $M[j+1]$ and
$B[j+1]$) onwards until it reaches the $i$-th level or one of them is marked
{\it invalid} in some level between $j$ and $i$. Both the situations are handled
in a manner similar to Cases (i) and (ii)  as stated above. 
Thus, each {\it valid} point in the $i$-th iteration has to
qualify as a {\it valid} point in the tests of all the $i-1$ levels. For
any other point the number of tests is at most $i-2$. This leads to the
following result:

\vspace{-0.1in}
\begin{lemma}\label{lx}
In the $i$-th iteration, the amortized time complexity of finding all valid
pairs is $O(ni)$.
\end{lemma}
\vspace{-0.1in}


The main task in the $i$-th iteration is to find the median of the slope of
lines
 corresponding to valid pair of points.
We essentially use the median finding algorithm of Munro and Raman
\cite{MunroR96} for this
purpose (see Appendix 1). Notice that, in order to get each slope, we need to
get a {\it valid} pair of points, which takes $O(i)$ time (see Lemma \ref{lx}).
The time required for finding the lowest slope is $O(ni)$. Similarly,
computing the second lowest needs another $O(ni)$ time. Proceeding similarly,
the time complexity of the procedure ${\cal A}_0$ of \cite{MunroR96} is
$O(n^2i^2)$ (see the Appendix). Similarly, ${\cal A}_1$ takes $O(i^2n^{1+
\frac{1}{2}} \log n )$
time, and so on. Finally, ${\cal A}_k$ takes $O(i^2
n^{(1+\frac{1}{k+1})}\log^{k} n)$ time.
Choosing $k = \sqrt{\frac{\log n}{\log \log n}} < \log n$, we need $O(\log n)$
space in total. Thus, we have the following result:

\vspace{-0.1in}
\begin{lemma} \label{ly}
The time complexity of the $i$-th iteration of the while-loop of the procedure \CB\ is $O(i^2
n^{(1+\frac{1}{k+1})}\log^{k} n)$, where $1 \leq k \leq \sqrt{\frac{\log
n}{\log \log n}}$. The extra space required is $O(\log n)$. 
\end{lemma}
\vspace{-0.1in}

At the end of $O(\log n)$ iterations, we could discard all the points
except at most $|IndexP|+3$ points, where $|IndexP|$ is the number of cells in
the array $IndexP$ that contain valid indices of $P$ ($\neq -1$). This can be at
most $O(\log n)$ in number. We can further prune the points in the $IndexP$
array using
the
in-place algorithm for \CB\ described in \cite{Vahrenhold12}.
Thus, we have the following result: 

\vspace{-0.1in}
\begin{lemma} \label{lv}
The read-only version of the procedure \CB\ is correct and
the  time complexity  is  
$O(n^{(1+\frac{1}{k+1})}\log^{k+3} n)$,
where where $1 \leq k \leq \sqrt{\frac{\log
n}{\log \log n}}$.  Apart from
the input array, it requires $O(\log n)$ extra space.
\end{lemma}
\vspace{-0.1in}
\begin{proof}
The correctness of this read-only version of the procedure \CB\ follows from
the fact that the Invariant~\ref{inv1} is correctly maintained.

By Lemma \ref{ly}, the time complexity of the $i$-th iteration is $O(i^2
n^{(1+\frac{1}{k+1})}\log^{k} n)$, where $i=1,2,\ldots,\log n$. Thus, the total
time complexity of all the $O(\log n)$ iterations is
$O(n^{(1+\frac{1}{k+1})}\log^{k+3} n)$. The time required by the in-place
algorithm for considering
all the entries in the array $IndexP$ is $O(\log n)$.

The space complexity obviously follows since the same set of arrays $M$,
$B$, $IndexP$ and the stack for finding the median can be used for all the
iterations, and each one is of size at most $O(\log n)$.  \qed
\end{proof}

\vspace{-0.1in}
 \subsubsection{Correctness and Complexity - Proof of Theorem \ref{th}}
\vspace{-0.1in}  
The correctness of the algorithm \ROUH\ follows from the correctness of
Kirkpatrick and Siedel's algorithm~\cite{KirkpatrickS86}, as we are following
the basic structure of this. The procedure \CB\ is evoked $h$ times, where $h$
is the number of hull-edges. Consider the recursion tree of the algorithm
\ROUH. Note that the depth of this tree is $O(\log n)$ (more specifically,
$\log h$), and total time complexity of a single level is
$O(n^{(1+\frac{1}{k+1})}\log^{k+3} n)$,  where $1 \leq k \leq
\sqrt{\frac{\log n}{\log \log n}}$ (see Lemma \ref{lv}). As there are at most
$\log n$
levels, so the total time complexity of the algorithm \ROUH\ is
$O(n^{(1+\frac{1}{k+1})}\log^{k+4} n)$. Substituting $\frac{\epsilon}{2}=
\frac{1}{k+1}$ and then $n^\frac{\epsilon}{2} \geq
\log^{4+\frac{1}{\epsilon}}n$, we have time complexity  $O(n^{1+\epsilon})$,
where 
$\epsilon$ satisfies  $\sqrt{\frac{\log\log n}{\log n}} <
\epsilon <1$.

For the recursion of \ROUH\ we need $\log n$ space and for each node of the
recursion tree we need another $O(\log n)$ space for the procedure \CB. However,
we can re-use the same space for each of the nodes for computing the bridge.
Hence the total space complexity of the algorithm \ROUH\ is
$O(\log n)$.

\vspace{-0.22in}
\section{2D Linear Programming}
\vspace{-0.17in}
In this section, we consider the problem of solving 2D linear programming in a read-only
setup, i.e, the constraints are given in a memory where swapping  of elements
or modifying any entry is not permissible. Megiddo proposed a linear time
prune-and-search algorithm for this problem which takes $O(n)$ space
~\cite{Megiddo83a}. We will show that Megiddo's algorithm for 2D linear
programming can be implemented when the constraints are stored in a read-only
memory using $O(\log n)$ extra-space and the running time would be
$O(n^{1+\epsilon})$, where $\sqrt{\frac{\log\log n}{\log n}} < \epsilon <1$.

\vspace{-0.2in}
\subsection{Overview of Megiddo's 2D Linear Programming}
\vspace{-0.15in}
The 2D linear programming problem is as follows:

\vspace{-0.1in}
\begin{tabbing}
subject to: \= xxxxxxxxxxxxxxxxx\= \kill
\>\>$\min_{x_1,x_2} c_1x_1+c_2x_2$\\
subject to: \>\>$a'_ix_1+b'_ix_2 \geq \beta_i$, 
$i \in
I=\{1,2, \ldots n\}$. 
\end{tabbing}
\vspace{-0.1in}

For ease of designing a linear time algorithm, Megiddo transformed it to 
an
equivalent form, stated below:

 \vspace{-0.1in}
\begin{tabbing}
subject to: \= xxxxxxxxxxxxxxxxx\= \kill
\>\>$\min_{x,y}  y$\\
subject to: \>\>$y \geq a_ix+b_i$, $i \in I_1$,\\
\>\>$y \leq a_ix+b_i $, $i \in I_2$,\\
\>\>$|I_1|+|I_2| \leq n$.
\end{tabbing}
\vspace{-0.1in}

 \begin{algorithm}
 \small
 \caption{\MDLP($I,c_1,c_2$)}
 \KwIn{A set of $n$ constraints $a'_ix_1+b'_ix_2 \geq \beta_i$, for $i \in
I=\{1,2, \ldots n\}$, }
\KwOut{The value of $x_1, x_2$ which minimizes $c_1x_1+c_2x_2$} 
*(Uses prune-and-search technique)*\\
{\bf STEP 1:} Convert the form into the following:
$\min_{x,y}  y$, subject to \\
(i) $y \geq a_ix+b_i $, $i \in I_1$, (ii) $y \leq a_ix+b_i $, $i \in I_2$, where
$|I_1|+|I_2| \leq n$.\\

{\bf STEP 2:}
Set $a=-\infty$ and $b=\infty$\;
\While{$|I_1 \cup I_2| > 4$}
{

{\bf STEP 2.1:} Arbitrarily pair-up the constraints of $I_1$ (resp. $I_2$).
Let $M_1$ (resp. $M_2$) be the set of aforesaid pairs, where
$|M_1|=\frac{|I_1|}{2}$ and  $|M_2|=\frac{|I_2|}{2}$.\\
Each pair of constraints in $M_1 \cup M_2$ are denoted by $(i,j)$ where $i$ and
$j$ indicate the $i$-th and $j$-th constraints. \\
  {\bf STEP 2.2:}\\
  \For{each pair $(i,j) \in M_1 \cup M_2$}
  {
  {\bf if} $a_i \neq a_j$ {\bf then}  Compute
$x_{ij}=\frac{b_i-b_j}{a_j-a_i}$\;
  }

  Find the median $x_m$ among all $x_{ij}$'s which are in the interval $[a,b]$
\;
  
  {\bf STEP 2.3:} Test  whether optimum $x^*$ satisfies $x^*=x_m$ or $x^*>x_m$
or $x^*<x_m$ as follows:\\
  {\bf STEP 2.3.1:} Compute $g=\max_{i\in I_1} a_ix_m+b_i$; $h=\min_{i\in I_2}
  a_ix_m+b_i$;\\
  {\bf STEP 2.3.2:} Compute\\ 
   \hspace{0.2in}$s_g=\min{a_i|i \in I_1, a_ix_m+b_i = g}$; 
   $S_g=\max{a_i|i \in I_1, a_ix_m+b_i = g}$; \\
   \hspace{0.2in}$s_h=\min{a_i|i \in I_2, a_ix_m+b_i = h}$; 
   $S_h=\max{a_i|i \in I_2, a_ix_m+b_i = h}$; \\
  {\bf STEP 2.3.2:}  (* $g \leq h$ $\Rightarrow$ $x_m$ is feasible ;
$g > h$ $\Rightarrow$ $x_m$ in infeasible region *)\\
  \If{$g > h$} 
  {{\bf if} $s_g > S_h$ {\bf then} (* $x_m < x^*$ *) $b=x_m$\\
  {\bf if} $S_g < s_h$ {\bf then} (* $x_m > x^*$ *) $a=x_m$\\
  {\bf else} Report there is no feasible solution of the LP problem; 
  {\bf Exit}}
  \Else  
  {{\bf if} $s_g > 0~ \& ~s_g \geq S_h$ {\bf then} (* $x_m < x^*$ *) $b=x_m$\\
  {\bf if} $S_g < 0 ~\& ~S_g \leq s_h$ {\bf then} (* $x_m > x^*$ *) $a=x_m$\\
  {\bf else} Report optimum solution $x_1=x_m$ \& $x_2=\frac{g-c_1x_1}{c_2}$;
{\bf Exit}}

  {\bf STEP 2.4:} *(Pruning step - 
   The case where iteration continues.\\ \hspace{0.7in} Without loss of
generality Assume that $x^*>x_m$;  *)\\
    \For{each pair $(i,j) \in M_1 \cup M_2$}
    {
    {\bf If} {$a_i=a_j$} {\bf then} Ignore one of the two constraints\;
    {\bf If} {$a_i \neq a_j$ and $x_{ij} <x_m$} {\bf then} Ignore one of the two
    constraints\;
    }

}
{\bf STEP 3:} *(The case when $|I_1 \cup I_2| \leq 4$)*\\
  The problem can be solved directly.
\normalsize
\label{MDLP}
\end{algorithm}

Megiddo's 2D linear programming algorithm uses prune-and-search
technique. It maintains an interval $[a,b]$ of feasible values of $x$ (i.e., $a
\leq x \leq b$). At the beginning of the algorithm, $a=-\infty$ and $b=\infty$.
After each iteration of the algorithm, either it finds out that at some $x=x_m$
($a\leq x_m \leq b$) the optimal solution exist (so the algorithm stops) or the 
interval $[a,b]$ is redefined (the new interval is either $[a,x_m]$ or
$[x_m,b]$) and  at least $\frac{n}{4}$ constraints are pruned for the next
iteration. The detail steps of the algorithm is given in the algorithm
\MDLP($I,c_1,c_2$).

Megiddo's 2D linear programming algorithm needs $O(n)$ time and $O(n)$ space.
In the next subsection we will show how to tailor this algorithm to work in the
read-only setup such that it does not take more than $O(\log n)$ space and
running time is $O(n^{1+\epsilon})$, where  $\sqrt{\frac{\log\log n}{\log n}} <
\epsilon <1$.

\vspace{-0.1in}
\subsection{2D-Linear Programming in Read-only setup}
\vspace{-0.1in}

We will give step by step description of implementing \MDLP\ in a read-only
setup. The straight-forward  conversion of one form into another mentioned in 
STEP 1 would take $O(n)$ extra-space. Note that  remembering only the objective
function 
$y=c_1x_1+c_2x_2$, will enable one to reformulate the newer version of the
constraints on-the-fly substituting $x_2$ in terms of $x_1$ and $y$ (replacing
$x_1$ by $x$) in each
constraint. So, we need not to worry about storing this new form.
 It is also to be
noted that 2D linear programming can be implemented in an in-place model in
$O(n)$ time using $O(1)$ extra-space~\cite{BronnimannIKMMT02}. So, we can
implement the pruning activities in Step 2 in read-only environment 
in a manner similar  to \CB\ as
described in section 2.2 using $O(\log n)$ space and
$O(n^{1+\epsilon})$ time,
where $\epsilon$ satisfies  $\sqrt{\frac{\log\log n}{\log n}} <
\epsilon <1$. Step 3 can obviously be implemented when the
constraints are given in a read-only memory.  
Hence, we have the following result:

\vspace{-0.1in}
\begin{theorem}
 2D linear programming can be implemented in a read-only model in
$O(n^{1+\epsilon})$ time  using $O(\log n)$ extra-space, where 
$\sqrt{\frac{\log\log n}{\log n}} <
\epsilon <1$. 
\end{theorem}

\vspace{-0.25in}
\section{3D Linear Programming}
\vspace{-0.1in}

In the same paper~\cite{Megiddo83a} Megiddo proposed a linear time algorithm for
3D linear programming. The problem is stated as follows:

\vspace{-0.1in}
\begin{tabbing}
subject to: \= xxxxxxxxxxxxxxxxx\= \kill
\>\>$\min_{x_1,x_2,x_3} d_1x_1+d_2x_2+d_3x_3$\\
subject to: \>\>$a'_ix_1+b'_ix_2+c'_ix_3 \geq \beta_i$, 
$i \in
I=\{1,2, \ldots n\}$. 
\end{tabbing}
\vspace{-0.1in}

As earlier, Megiddo transformed the problem into the following
equivalent
form:

\vspace{-0.1in} 
\begin{tabbing}
subject to: \= xxxxxxxxxxxxxxxxx\= \kill
\>\>$\min_{x,y,z}  z$\\
subject to: \>\>$z \geq a_ix+b_iy+c_i$, $i \in I_1$,\\
\>\>$z \leq a_ix+b_iy+c_i $, $i \in I_2$,\\
\>\>$0 \geq a_ix+b_iy+c_i $, $i \in I_3$,\\
\>\>$|I_1|+|I_2|+|I_3| \leq n$.
\end{tabbing}
\vspace{-0.1in}

\MDLPT\  algorithm also follows prune-and-search paradigm. It
pairs-up constraints $(C_k^i, C_k^j)$ where $C_k^i, C_k^j$
are from same set $I_k, k\in\{1,2,3\}$; So, there are at most $\frac{n}{2}$
pairs. Let $C_k^i$ (resp. $C_k^j$) corresponds to $a_ix+b_iy+c_i$ (resp. 
$a_ix+b_iy+c_i$). If $(a_i,b_i) = (a_j,b_j)$, then we can easily ignore one of
the constraints. Otherwise (i.e., if $(a_i,b_i) \neq (a_j,b_j)$), then each pair
signifies a line $L_{ij}$: $a_ix+b_iy+c_i = a_jx+b_jy+c_j$ which divides the
plane into two halves. Let $\cal L$ be the set of lines obtained in this way.
We compute the median $\mu$ of the gradients of the members in $\cal L$. Next,
we pair-up the members in $\cal L$ such that one of them have gradient
less than $\mu$ and the other one have gradient greater than $\mu$. Let $\Pi$
be the set of these paired lines. Each of these pairs will intersect. We compute
the intersection point $a$ having median $y_m$ among the $y$-coordinates of
these intersection points. Next, we execute the procedure \TL\ as stated below
with respect to the line $L_H: y=\mu x + y_m\sqrt{\mu^2+1}$ (having gradient
$\mu$ and passing through $a$). This determines in which side of $L_H$ the
optimum solution lies. Next, we identify the pairs  in $\Pi$ which intersect on
the other side of the optimum solution. Among these pairs, we compute the
intersection point $b$ having median $x$-coordinates of their intersections, and
execute \TL\ with the line $L_V$ having gradient $\frac{1}{\mu}$ and passing
through  $b$. $L_H$ and $L_V$ determines a quadrant $Q$ containing the optimum
solution. Now consider the paired lines in $\Pi$ that intersect in the quadrant
$Q'$, diagonally opposite to $Q$. Let $(L_{ij}, L_{k\ell})$ be a paired line of
$\Pi$ that intersect in $Q'$. For at least one of the lines $L_{ij}$ and
$L_{k\ell}$, it is possible to correctly identify the side containing the
optimum solution without executing \TL\ (see \cite{Megiddo83a}). Thus, for each
of such lines we can prune one constraint. 
As a result, after each iteration  it can prune at-least $\frac{n}{16}$
constraints for
next iteration or report optimum.

The procedure \TL\ takes a straight line $L$ in the $x$-$y$ plane and
tests whether the solution on the line $L$ (i) does not exist, or (ii)
unbounded, or (iii) unique optimum solution exists using a 2D linear
programming. In Case (ii), the solution of the given 3D linear programming
problem is unbounded. In Case (i) and (iii), we need to decide in which side of
$L$ the optimum solution of the given linear programming problem lies by
executing two other 2D linear programming. Both of them can be executed if the
constraints are given in read-only memory.

 The detail description of the algorithm
\MDLPT\ is given in Appendix 2. As \MDLPT\ is a {\it prune-and-search}
algorithm,
one can easily show that this can be implemented in a read-only setup. Thus, we
have the following result:

\vspace{-0.2in}
\begin{theorem}
3D linear programming can be implemented in a read-only model in
$O(n^{1+\epsilon})$ time  using $O(\log n)$ extra-space, where 
$\sqrt{\frac{\log\log n}{\log n}} <
\epsilon <1$. 
\end{theorem}
\remove{
\section{Conclusion}
In this paper, we considered three important problems in geometric optimization
that can be solved when the input is given in a read-only array. These are 
a restricted variation of the convex hull of a 2D point set where the points
are sorted with respect to their $x$-coordinates, and linear programming in two
and three dimension. The time complexities of all these algorithms are
$O(n^{1+\epsilon})$ and each of then uses $O(\log n)$ extra work-space. To the
best of our knowledge this is the first attempt for solving these problems
deterministically in read-only setup with sub-quadratic time using sub-linear
space. It would be a nice challange to solve any one of these
problems deterministically in the same model with sub-quadratic time using
$O(1)$ extra space. 
}

\newpage
\centerline{Appendix 1} 

{\bf Munro and Raman's median finding algorithm:}
Given a set of $n$ real numbers in a read-only array $P$, the
algorithm in \cite{MunroR96} for finding their median  
 is designed by using a set of procedures  ${\cal A}_0,{\cal
A}_1, {\cal A}_2,\ldots, {\cal A}_k $, where procedure ${\cal A}_i$ finds the
median by evoking the procedure ${\cal A}_{i-1}$ for $i \in \{1,2,\ldots, k\}$.
The procedures ${\cal A}_0,{\cal A}_1, {\cal A}_2,\ldots, {\cal A}_k$ are
stated below.

{\bf Procedure ${\cal A}_0$:} In the first iteration, after checking all the
elements in $P$, it finds the largest element $p_{(1)}$ in linear time. In the 
second iteration it finds the second largest $p_{(2)}$ by checking only the
elements which are less than $p_{(1)}$. Proceeding in this way, in the $j$-th
iteration it finds the $j$-th largest element $p_{(j)}$ considering all the
elements in $P$ that are less than $p_{(j-1)}$. In order to get the median we
need to proceed up to $j=\lfloor\frac{n}{2}\rfloor$. Thus, this simple median
finding algorithm takes $O(n^2)$ time and $O(1)$ extra-space.

{\bf Procedure ${\cal A}_1$:} It divides the array $P$ into blocks of size
$\sqrt{n}$ and in each block it finds the median using Procedure ${\cal A}_0$.
After computing the median $m$ of a block, it counts the number of elements in
$P$ that are smaller than $m$, denoted by $\rho(m)$, by checking all the
elements in the array $P$. It maintains two best block medians $m_1$ and $m_2$,
where $\rho(m_1) = \max\{\rho(m)|\rho(m) \leq \frac{n}{2}\}$, and $\rho(m_2) =
\min\{\rho(m)|\rho(m) \geq \frac{n}{2}\}$. Thus, this iteration needs
$O(n\sqrt{n})$ time. 

After this iteration, all the elements $P[i]$ satisfying $P[i] < m_1$ or $P[i] >
m_2$ are marked as {\it invalid}. This does not need any mark bit; only one
needs to remember $m_1$ and $m_2$. In the next iteration we again consider same
set of blocks, and compute the median ignoring the {\it invalid} elements. 

Since, in each iteration $\frac{1}{4}$ fraction of the existing {\it valid}
elements are marked {\it invalid}, we need at most $O(\log n)$ iterations to
find the median $\mu$. Thus the time complexity of this procedure is
$O(n\sqrt{n} \log n)$.

{\bf Procedure ${\cal A}_2$:} It  divides the whole array into $n^{1/3}$ blocks
each of size $n^{2/3}$, and computes the block median using the procedure ${\cal
A}_1$. Thus, the overall time complexity of this procedure for computing the
median is $n^{1+\frac{1}{3}} \log ^2 n$. 

Proceeding in this way, the time complexity of the procedure ${\cal A}_k$ will
be $O(n^{(1+\frac{1}{k+1})}\log^{k} n)$. As it needs a stack of depth $k$  for
the recursive evoking of ${\cal A}_{k-1}$, ${\cal A}_{k-2}$, $\ldots$, ${\cal
A}_0$, the space complexity of this algorithm is $O(k)$. 

Setting $\epsilon=\frac{1}{k+1}$, gives the running time as 
$O(\frac{n^{1+\epsilon}\log^{\frac{1}{\epsilon}}n}{\log n})$. If we choose
$n^\epsilon=\log^{\frac{1}{\epsilon}}n$, then  $\epsilon$ will be
$\sqrt{\frac{\log\log n}{\log n}}$, and this will give the running time 
$O(\frac{n^{1+2\epsilon}}{\log n})$, which is of $O(n^{1+2\epsilon})$. 
So, the general result is as follows:

\begin{result}
 \label{res1}
For a set of $n$ points in $\IR$ given in a read-only memory, the median  can be
found in $O(n^{1+\epsilon})$ time with $O(\frac{1}{\epsilon})$ extra-space,
where $2\sqrt{\frac{\log \log n}{\log n}} \leq \epsilon < 1$.
\end{result}

 \begin{algorithm}[h]
 \small
 \caption{Appendix 2 - \MDLPT($I,c_1,c_2$)}
 \KwIn{A set of $n$ constraints $a'_ix_1+b'_ix_2+c'_ix_3 \geq \beta_i$, for $i
\in I=\{1,2, \ldots n\}$, }
\KwOut{The value of $x_1, x_2$ which minimizes $c_1x_1+c_2x_2$} 
(* Uses prune-and-search technique *)\\
{\bf STEP 1:} Convert the form into  following: 
$\min_{x,y,z}  z$, subject to 
(i) $y \geq a_ix+b_iy+c_i $, $i \in I_1$, (ii) $y \leq a_ix+b_iy+c_i $, $i \in
I_2$, (iii)
$0 \geq a_ix+b_iy+c_i$, $i \in I_3$, where
$|I_1|+|I_2|+|I_3| \leq n$.\\

{\bf STEP 2:}\\
\While{$|I_1 \cup I_2 \cup I_3| \geq 16$}
{

  {\bf STEP 2.1:} Arbitrarily pair-up the constraints $a_ix+b_iy+c_i$,
$a_jx+b_jy+c_j$ where $i,j$ are from same set $I_k,
k\in\{1,2,3\}$.
Let   ${\cal L}_1$, ${\cal L}_2$ and ${\cal L}_3$ be the set of aforesaid pairs
and ${\cal L}={\cal L}_1 \cup {\cal L}_2\cup {\cal L}_3$. \\

  {\bf STEP 2.2:}
 Let ${\cal L}_C=\{(i,j)\in {\cal L}| (a_i,b_i) \neq (a_j,b_j) \}$ and
  ${\cal L}_P=\{(i,j)\in {\cal L}| (a_i,b_i)= (a_j,b_j) \}$\;
  Compute the median $\mu$ of the slopes $\alpha(L_{ij)}$ of all the straight
lines $L_{ij}$:
$a_ix+b_iy+c_i=a_jx+b_jy+c_j$, $(i,j) \in {\cal L}_C$\;

 {\bf STEP 2.3:}\\

 Arbitrarily pair up $(L_{ij}, L_{i'j'})$ where $\alpha(L_{ij}) \leq \mu \leq
\alpha(L_{i'j'})$ 
and $(i,j), (i',j') \in {\cal L}_C$. Let $M$ be the set of these
$\lfloor\frac{n}{4}\rfloor$ pairs of lines\; 
Let $M_P=\{(L_i,L_j) \in M|\alpha(L_i)=\alpha(L_j)=\mu\}$ (* parallel line-pairs
*) and 
$M_I=\{(L_i,L_j)\in M|\alpha(L_i)\neq \alpha(L_j)\}$ (* intersecting line-pairs
*)\;

  \For{each pair $(L_i,L_j) \in M_P$} {compute $y_{ij}=\frac{d_i+d_j}{2}$, where
$d_i$ = distance of $L_i$ from the
line $y=\mu x$}
\For{each pair $(L_i,L_j) \in M_I$}{Let $a_{ij}$ = point of intersection of 
$L_i$ \& $L_j$, and $b_{ij}$ = projection of $a_{ij}$ on
$y=\mu x$. Compute 
$y_{ij}$ = signed distance of the pair of points $(a_{ij},b_{ij})$, and
$x_{ij}$ = signed distance of $b_{ij}$ from the origin\;
}
Next, compute the median $y_m$ of the $y_{ij}$ values corresponding to all the 
pairs in $M$\;
{\bf Step 2.4:} Consider the line $L_H: y=\mu x+y_m\sqrt{\mu^2+1}$, which 
is parallel to $y=\mu x$ and at a distance $y_m$ from  $y=\mu x$\;

Test on which half-plane defined by the line $L_H$ contains the optimum by
evoking \TL($L_H$)
  
 {\bf Step 2.5:} Let $M_I'$ = \{$(L_i,L_j) \in M_I|$   
$a_{ij}$ \& $\pi^*$ lie in  different sides  of  $L_H$\}\;

Compute the median $x_m$ of $x_{ij}$-values for the line-pairs in $M_I'$.
Define a line $ L_V$ perpendicular  to $y=\mu x$ and passing through a
point on $y=\mu x$ at a distance $x_m$ from the origin\;  
Execute the procedure \TL($L_V$)  and decide in
which
side of $ L_V$ the optimum lies\;
We consider $ L_H$ and
$ L_V$ as horizontal and vertical lines respectively; W.L.O.G., assume that
optimum lies in the top-left quadrant\;

 {\bf Step 2.6:} (* Pruning step *)\\ 
\For{all the members $(L_i,L_j) \in M_I$ whose points of intersection
($a_{ij}$) lie in the bottom-right quadrant}
{
Discard one of the four constraints defined by the pair of lines $L_i,L_j$  

}

\For{all the members $(L_i,L_j) \in M_P$ whose $y_{ij} \leq y_m$}
{
 Discard one of the four constraints defined by the pair of lines $L_i,L_j$ 

}

}
{\bf STEP 3:} *(The case when $|I_1 \cup I_2 \cup I_3| \leq 16$)*\\
  The problem can be solved directly by brute-force manner.

\normalsize
\label{MDLPT}
\end{algorithm}

\end{document}